\documentclass[aps,preprint,superscriptaddress,floatfix,preprintnumbers,nofootinbib,showpacs,11pt,a4paper]{revtex4-1}
\usepackage[utf8]{inputenc}

\pdfoutput=1

\usepackage[english]{babel}
\usepackage{amsmath,amssymb,amsfonts}
\usepackage{graphicx}
\usepackage{booktabs}
\usepackage{hyperref}
\usepackage{color,xcolor}
\usepackage[sort&compress]{natbib}
\usepackage{tikz}
\usepackage{slashed}
\usepackage{url}
\usepackage[makeroom]{cancel}
\usepackage{geometry}

\def\beqn{\begin{eqnarray}}
\def\eeqn{\end{eqnarray}}
\def\beqs{\begin{subequations}}
\def\eeqs{\end{subequations}}
\def\beq{\begin{equation}}
\def\eeq{\end{equation}}
\def\ba{\begin{array}}
\def\ea{\end{array}}

\def\non{\nonumber\\}

\def\[{\left[}
\def\]{\right]}
\def\({\left(}
\def\){\right)}

\def\GeV{\rm GeV}

\def\gSU{\rm SU}

\newcommand{\rep}[1]{\mathbf{#1}}
\newcommand{\repb}[1]{\mathbf{\overline{#1}}}

\def\mG{\mathcal{G}}

\def\mL{\mathcal{L}}

\def\mO{\mathcal{O}}

\def\mR{\mathcal{R}}

\begin{document}
\title{\Large High-quality grand unified theories \\ with three generations}
\bigskip

\author{Ning Chen}
\email[ ]{chenning$\_$symmetry@nankai.edu.cn}
\affiliation{ School of Physics, Nankai University, Tianjin, 300071, China}
%

\begin{abstract} 
We extend the unitary groups beyond the ${\rm SU}(5)$ and ${\rm SU}(6)$ to look for possible grand unified theories that give rise to three-generational Standard Model fermions without the simple repetition. 
By demanding asymptotic free theories at short distances, we find gauge groups of ${\rm SU}(7)$, ${\rm SU}(8)$ and ${\rm SU}(9)$ together with their anomaly-free irreducible representations are such candidates.
Two additional gauge groups of ${\rm SU}(10)$ and ${\rm SU}(11)$ can also achieve the generational structure without asymptotic freedom.
We also find these models can solve the Peccei-Quinn (PQ) quality problem which is intrinsic in the axion models, with the leading PQ-breaking operators determined from the symmetry requirement.
\end{abstract}
\maketitle

\baselineskip=16pt
\pagenumbering{arabic}

\vspace{1.0cm}

\section{Introduction}
%
%
Grand Unified Theories (GUTs)~\cite{Georgi:1974sy,Fritzsch:1974nn} were proposed to unify all fundamental interactions described by the Standard Model (SM).
Aside from the aesthetic aspect of achieving the gauge coupling unification in its supersymmetric (SUSY) extension~\cite{Dimopoulos:1981zb}, it is pragmatic to conjecture the zeroth law of GUT, namely, {\it a successful GUT could address all intrinsic SM puzzles and as many physical issues beyond the SM as possible, with all necessary but the minimal set of fields determined by symmetry.}
One such longstanding puzzle that has not been well answered is the existence of three generational SM fermions, as well as their mass hierarchies in the framework of GUTs~\footnote{Some of the previous efforts in addressing the SM fermion masses in GUTs include Refs.~\cite{Georgi:1979df,Barr:1979xt,Georgi:1979ga,Georgi:1980sb,Dimopoulos:1991yz,Giudice:1992an,Ramond:1993kv,Hall:1993gn,Ibanez:1994ig,Nath:1996qs,Barr:2007ma}. }.
In a seminal paper~\cite{Georgi:1979md}, Georgi suggested to extend the minimal ${\rm SU}(5)$ into larger simple Lie group ${\rm SU}(N)$ (with $N\geq 7$), and built his three laws of GUTs.
Instead of simple repetition of a set of anomaly-free irreducible representations (irreps) for three times, it is argued that the three-generational structure arises from different anti-symmetric irreps of the ${\rm SU}(N)$.

In this paper, we investigate the possible non-minimal GUTs beyond the ${\rm SU}(5)$ and ${\rm SU}(6)$ that can give rise to three-generational SM fermions.
The number of generations can be easily obtained according to the counting method in terms of the ${\rm SU}(5)$ irreps as given in Ref.~\cite{Georgi:1979md}.
It turns out that the SM fermion generations $n_g$ can already become three or beyond for the ${\rm SU}(7)$ group already~\cite{Frampton:1979cw}.
Historically, the number of the SM fermion generations $n_g$ was also considered to be beyond three~\cite{Frampton:1999xi}.
Meanwhile, the direct searches for the fourth-generational quarks at the Large Hadron Collider (LHC) have already excluded this possibility~\cite{ATLAS:2012tkh,ATLAS:2012aw,CMS:2012jki}.
Therefore, only the non-minimal GUTs with their anomaly-free irreps that lead to $n_g=3$ cases will be considered in our study.
In Georgi's third law, he decided that no individual irrep of the GUT group should appear more than once.
Accordingly, he found that the minimal GUT group that give rise to $n_g=3$ is ${\rm SU}(11)$, with a total number of $1,023$ left-handed fermions~\cite{Georgi:1979md}. 
Obviously, the third law prevent the three-generational structure through the simple repetition of the anomaly-free irreps.
However, this may be a too strong constraint and was not usually adopted in the later studies.
In our discussions, we modify Georgi's third law in a different version proposed by Christensen and Shrock~\cite{Christensen:2005bt} in the study of the dynamical origin of the SM fermion masses.
A different point of view can be made such that the global symmetries can usually emerge once the original third law was abandoned, such as in the ${\rm SU}(9)$ GUT~\cite{Georgi:1981pu}. 
This can be advantageous at least in two aspects.
The first advantage is the emergent global symmetry, with its breaking, can be a mechanism to explain the lightness of the Higgs boson, as was discussed by Dvali~\cite{Dvali:1993yf} in the context of the SUSY ${\rm SU}(6)$.
The other advantage is that the global ${\rm U}(1)$ symmetry can be identified as the Peccei-Quinn (PQ) symmetry~\cite{Peccei:1977hh} for the strong CP problem.
The emergent PQ symmetry, together with both the gauge and the global symmetries, can usually constrain the mass dimensions of the PQ-breaking operators and lead to a high-quality axion~\cite{Georgi:1981pu,Dine:1986bg,Barr:1992qq,Kamionkowski:1992mf,Holman:1992us,Ghigna:1992iv}.
Two recent examples include the axion from the ${\rm SO}(10)$~\cite{DiLuzio:2020qio} and the SUSY ${\rm SU}(6)$ GUT~\cite{Chen:2021haa}.

The rest of the paper is organized as follows.
In Sec.~\ref{sec:remark}, we review Georgi's guidelines of building the non-minimal GUTs that can lead to three generations of SM fermions without simple repetition.
Some other relevant results of the gauge anomaly cancellation, the Higgs representations, and the PQ quality are also setup there.
Sec.~\ref{sec:results} is the core of this work.
We analyze all possible ${\rm SU}(N)$ GUTs (up to ${\rm SU}(11)$) and their anomaly-free fermion contents that can lead to three generational SM fermions according to Georgi's counting.
The PQ charge assignments to Higgs fields and the corresponding PQ-breaking operators will be presented.
We summarize our results and make discussions in Sec.~\ref{sec:conclusion}.

\section{Some general remarks}
\label{sec:remark}
%
%

\subsection{Lie group representations and Georgi's guidelines}

To facilitate the discussion, we express the fermion representations under the ${\rm SU}(N)$ GUT group in terms of the set of rank-$k$ anti-symmetric irrep of $[N\,,k]$ as follows
\beqn\label{eq:GUTfermion}
\{ f_L \}_{ {\rm SU}(N) }&=& \sum_{ k=0}^{N-1}\, n_k\, [ N\,,k]\,,
\eeqn
with $n_k$ being the multiplicity.
Obviously, $k=0$ corresponds to the singlet representation, and $k=1$ corresponds to the fundamental representation, and etc.
The singlet representations contribute neither to the gauge anomaly, nor to the renormalization group equations (RGEs).
Through out the discussion, we always denote the conjugate representation such that $\overline{ [N\,, k] } = [N\,, N-k] $.
It will be also useful to use a compact vector notation of
\beqn
\vec n &\equiv& ( n_0\,, ...\,, n_{N-1} ) \,.
\eeqn
For a given rank-$k$ anti-symmetric irrep of $[N\,,k]$, its dimension and trace invariants are
\beqn
{\rm dim} (  [ N\,,k]) &=& \frac{N!}{ k! (N-k)!} \,,\\
T([ N\,,k])&=& \frac{ (N-2)!}{ 2 (k-1)! (N-k-1)!} \,.
\eeqn

From Cartan's classification, it is well-known that the only possible Lie groups for non-minimal GUTs beyond the ${\rm SU}(5)$ or ${\rm SO}(10)$ are
\beqn
&& {\rm SU}(N) ~~ (N\geq 6)\,, \quad {\rm SO}(4k+2)~~ (k\geq 3)\,,\quad E_6\,.
\eeqn
Since the exceptional group of $E_6$ has fixed rank, it is impossible to consider further extensions.
For any irrep under these Lie groups, one can always decompose it under the subgroup of the ${\rm SU}(5)$.
For example, the fundamental representation of the ${\rm SU}(N)$ can be decomposed as
\beqn
 [ N\,,1]&=& (N-5)\times [5\,,0] \oplus [5\,,1] \,.
\eeqn
The decompositions of the higher irreps can be obtained by {\tt LieART}~\cite{Feger:2019tvk}.
For an ${\rm SU}(N)$ GUT, its fermion contents can be generally decomposed in terms of the ${\rm SU}(5)$ irreps as follows
\beqn
\{ f_L \}_{ {\rm SU}(N) }&=& n_0\, [ 5\,,0] + n_1\, [ 5\,,1] + n_2\, [ 5\,,2] + n_3\, [ 5\,,3] + n_4\, [ 5\,,4] \,.
\eeqn
The anomaly cancellation condition leads to the following relation to the multiplicities
\beqn
&& n_1+n_2=n_3 + n_4  \,.
\eeqn
In Ref.~\cite{Georgi:1979md}, Georgi argued that the counting of the SM fermion generations is equivalent to the counting of the multiplicity of the residual ${\rm SU}(5)$ irreps of $[5\,,2] \oplus [5\,,4]$, which is
\beqn\label{eq:Georgi_count}
&& n_g = n_2 - n_3 = n_4 - n_1 \,.
\eeqn
Note that the counting of the SM fermion generations in Eq.~\eqref{eq:Georgi_count} does not rely on the realistic gauge symmetry breaking patterns.
Based on Georgi's counting, it turned out that any GUT with orthogonal groups larger than the ${\rm SO}(10)$ essentially leads to $n_g=0$.
This can be understood by decomposing the $16$-dimensional ${\rm SO}(10)$ Weyl fermions under the ${\rm SU}(5)$ as
\beqn
\rep{16_F}&=& \rep{1_F} \oplus \repb{5_F} \oplus \rep{10_F}\,.
\eeqn
The Weyl fermions from larger orthogonal groups are always decomposed under the ${\rm SO}(10)$ in pairs of $\rep{16_F} \oplus \repb{16_F}$, and this can only lead to $n_g=0$.

Georgi's third law requires that not any representation of $[N\,,k]$ should appear more than once, which means $n_k=0$ or $n_k=1$ in Eq.~\eqref{eq:GUTfermion}.
This leads to a consequence that no global symmetry can emerge from the corresponding fermion setup.  
Instead, we adopt an alternative criterion by Christensen and Shrock~\cite{Christensen:2005bt}, namely, the greatest common divisor of $\{ n_k\}$ is not greater than unity.
Therefore, one can expect the global symmetry of
\beqn\label{eq:DRS}
\mG_{\rm global}&=& \prod_{ \{ k^\prime \} } \Big[ {\rm SU} ( n_{k^\prime} ) \otimes {\rm U}(1)_{k^\prime} \Big] \,,
\eeqn
for all irreps of $[N\,,k^\prime]$ that appear more than once.
This can be viewed as a generalization of the global symmetry in the rank-$2$ anti-symmetric theory of ${\rm SU}(N+4)$ by Dimopoulos, Raby, and Susskind (DRS)~\cite{Dimopoulos:1980hn} .
The ${\rm U}(1)$ components of the global symmetry~\eqref{eq:DRS} can be identified as the global PQ symmetry, which are likely to lead to high-quality axion~\cite{Chen:2021haa}.
In this regard, the modified criterion of the fermion assignments is likely to solve the long-standing PQ quality problem~\cite{Georgi:1981pu,Dine:1986bg,Barr:1992qq,Kamionkowski:1992mf,Holman:1992us,Ghigna:1992iv} in the framework of GUT.

\subsection{Gauge anomaly cancellation}

To have an anomaly-free non-minimal GUT, we have to solve the following Diophantine equation
\beqn\label{eq:anomaly_free}
&& \vec n \cdot \vec A =0\,,
\eeqn
with the $N$-dimensional anomaly vector~\cite{Banks:1976yg,Okubo:1977sc} being
\beqn
&& \vec A = \Big( A([N\,,0]) \,, A([N\,,1]) \,, ...\,, A([N\,,N-1]) \Big) \,,\label{eq:anomaly_vector}\\
&& A([N\,,0])= 0\,,\quad A([N\,,k])= \frac{ (N-2k) (N-3)! }{ (N-k-1)! (k-1)! }\,~(k> 0) \,.\label{eq:rk_anomaly}
\eeqn
The property that the anomaly of a given irrep and its conjugate cancel each other is apparent in Eq.~\eqref{eq:rk_anomaly}.
Also, the self-conjugate representations must be anomaly-free such that $A( [N\,, \frac{N}{2}])=0$ for $N$ being even.
Thus, the anomaly vector can be expressed as
\beqn
&& \vec A = \Big(0\,, A([N\,,1])\,, ...\,, A([N\,,[\frac{N}{2}] ])\,, - A([N\,,[\frac{N}{2}] ])\,, ... \,, -A([N\,,1])  \Big) \,.
\eeqn
In practice, one has to decompose the ${\rm SU}(N)$ fermion representations from $[N\,,1]$ to $\[ N\,,[ \frac{N}{2} ] \]$ under the ${\rm SU}(5)$ in order to count the generations.

\subsection{The Higgs representations}

Once the fermion contents are determined for a particular non-minimal GUT, the Higgs fields can be determined by the following criteria
\begin{enumerate}

\item The GUT symmetry breaking of ${\rm SU}(N) \to {\rm SU}(m) \otimes {\rm SU}(N- m) \otimes {\rm U}(1)$ with $m=[ \frac{N}{2} ]$ is always assumed at its first stage, which requires an adjoint Higgs field~\cite{Li:1973mq}.
The other possible symmetry breaking of ${\rm SU}(N) \to {\rm SU}(N-1)$ (with $N\geq 6$) at the first stage is very likely to lower the proton lifetime predictions, and thus bring tension with the current experimental constraint to the proton lifetime from the Super-Kamionkande~\cite{Super-Kamiokande:2020wjk}.

\item All possible gauge-invariant Yukawa couplings, which also respect the global symmetry in Eq.~\eqref{eq:DRS}, can be formed.

\item Higgs fields to achieve any intermediate symmetry breaking stages are necessary, where their proper irreps contain the SM-singlet directions.

\item Only the Higgs fields with the minimal dimensions are taken into account.

\end{enumerate}

Before proceeding to the more realistic models, we display the Higgs fields in the ${\rm SU}(6)$ GUT as an example.
Its minimal anomaly-free fermion contents and decomposition under the ${\rm SU}(5)$ are
\beqn
\{ f_L \}_{ {\rm SU}(6) }&=& [6\,,2] \oplus 2 \times [6\,,5] = \rep{15_F} \oplus 2\times \repb{6_F} \non
&=& 2\times [ 5\,,0] \oplus  [ 5\,,1] \oplus  [ 5\,,2]  \oplus 2\times [ 5\,,4] \,.
\eeqn
The $[ 5\,,1]$, one of the $[ 5\,,4]$, as well as two singlets of $[ 5\,,0]$ obtain their masses at an intermediate symmetry-breaking scale.
The remaining fermions contain precisely one generational SM fermions of $[ 5\,,2]  \oplus  [ 5\,,4] \approx [ 5\,,2]  \oplus  \overline{ [ 5\,,1]}$.
Apparently, the minimal ${\rm SU}(6)$ GUT has a global symmetry of ${\rm SU}(2)_{\repb{6_F} } \otimes {\rm U}(1)_{\rm PQ}$ and is a one-generational model according to Georgi's counting.
The gauge-invariant and ${\rm SU}(2)_{\repb{6_F} }$-invariant Yukawa coupling can be expressed as follows
\beqn
-\mL_Y&=& \repb{ 6_F}^\rho \rep{15_F} \repb{ 6_{H}}_\rho  + \rep{15_F} \rep{15_F} \rep{15_H} + \epsilon_{\rho\sigma } \repb{ 6_F}^\rho \repb{ 6_F}^\sigma ( \rep{15_H} + \rep{21_H} ) + H.c. \,.
\eeqn
with the minimal set of Higgs fields.
Of course, a $35$-dimensional adjoint Higgs field is necessary to achieve the first-stage GUT symmetry breaking of ${\rm SU}(6)\to {\rm SU}(3)_c \otimes {\rm SU}(3)_W \otimes {\rm U}(1)_X$~\cite{Li:2019qxy,Chen:2021haa,Chen:2021zwn}.
It turns out that the VEVs from Higgs fields of $( \rep{1}\,, \repb{3}\,, -\frac{1}{3} )_{ \mathbf{H}\,, \rho} \subset \repb{ 6_{H}}_\rho$ and $( \rep{1}\,, \rep{6}\,, +\frac{2}{3} )_{ \mathbf{H}}\subset \rep{21_H}$ can be responsible for the intermediate symmetry breaking of ${\rm SU}(3)_c \otimes {\rm SU}(3)_W \otimes {\rm U}(1)_X \to {\rm SU}(3)_c \otimes {\rm SU}(2)_W \otimes {\rm U}(1)_Y$~\cite{Chen:2021haa}.

\subsection{The asymptotic freedom (AF)}

The GUTs with their earliest versions are usually asymptotically free above the unification scale.
However, there was no definite answer whether the AF should be retained.
An alternative criterion is to have an asymptotic safe theory, which reaches a fixed point at the short distance~\cite{Litim:2014uca,Litim:2015iea}.
In general, the analysis of the asymptotic safe theories involves the RGEs of gauge couplings, as well as Yukawa and Higgs self couplings.
This can only be performed for individual theory by specifying the symmetry breaking patterns.
In the non-minimal GUTs, the AF is likely to be violated since the trace invariants of the rank-$2$ and rank-$3$ anti-symmetric representations scale as $T([N\,,2]) \sim N$ and $T( [N\,,3]) \sim N^2$.
Previously, this was also considered in the ${\rm SU}(11)$ model~\cite{Kim:1979uh}, but with only fermions taken into account.
In our discussions below, we study the short-distance behavior for non-minimal GUTs up to ${\rm SU}(11)$, with their minimal fermion setup.
It turns out the minimal models in ${\rm SU}(10)$ and ${\rm SU}(11)$ violate the AF, and thus careful analysis of their unification couplings and scales are necessary for these two cases.
The one-loop $\beta$ coefficients are obtained by including both fermions and Higgs fields as follows
\beqn\label{eq:b1}
b_1 &=& - \frac{11}{3} C_2(\mG ) + \frac{4}{3} \kappa \sum_f T(\mR_f ) + \frac{1}{3} \eta \sum_s T( \mR_s) \,, 
\eeqn
with $\kappa=1\,(1/2)$ for Dirac (Weyl) fermions, and $\eta=1\,(1/2)$ for complex (real) scalars.
For the adjoint Higgs fields, we always consider them to be real for the non-SUSY case.
The AF can be determined by whether $b_1<0$ or not.

\subsection{The PQ quality and axion}

The global PQ symmetry has an intrinsic problem known as the PQ quality~\cite{Georgi:1981pu,Dine:1986bg,Barr:1992qq,Kamionkowski:1992mf,Holman:1992us,Ghigna:1992iv}.
In general, global symmetries are not fundamental but arise with the underlying gauge theories.
They are believed to be broken by quantum gravity effects in the form of the following dimension-$2m+n$ operator
\beqn\label{eq:PQbreak_Op}
 \mO_{ \bcancel{\rm PQ} }^{d=2m+n} &=& k \frac{ |\Phi|^{2m} \Phi^n }{ M_{\rm pl}^{2m+n-4} } \,.
\eeqn
The size of PQ-breaking is constrained such that the minima of the QCD effective potential induced by axion should satisfy $ \Big| \langle a/f_a \rangle \Big| \lesssim 10^{-10}$, which leads to a PQ quality constraint of
\beqn\label{eq:PQquality}
&& \frac{f_a^d}{ M_{\rm pl}^{d-4} }  \lesssim 10^{-10}\, \Lambda_{\rm QCD}^4\,.
\eeqn
It turns out that the mass dimension in Eq.~\eqref{eq:PQbreak_Op} should be $d\gtrsim 9$ in order to have a reasonable axion decay constant $f_a\sim {\cal O}(10^{12})\,\GeV$ without much fine tuning of the coefficient $k$~\cite{Barr:1992qq,Kamionkowski:1992mf,Holman:1992us} in Eq.~\eqref{eq:PQbreak_Op}.
Without knowing underlying symmetry origin of the $\Phi$ field, there is generally no reason to forbid any PQ-breaking operators with $d\lesssim 9$.

Previous studies of the axion in the GUT~\cite{Wise:1981ry,DiLuzio:2018gqe,FileviezPerez:2019fku,FileviezPerez:2019ssf} were made in both ${\rm SU}(5)$ and ${\rm SO}(10)$, where the global PQ symmetry was introduced by hand. 
Therefore, the issue of PQ quality was still present.
Recent discussions of the PQ quality problem in the frame of GUT include the ${\rm SO}(10)$~\cite{DiLuzio:2020qio} and ${\rm SU}(6)$~\cite{Chen:2021haa} cases.
In the ${\rm SO}(10)$ GUT, the author made use of the generational symmetry in the limit of vanishing Yukawa couplings.
A dimension-$9$ gauge-invariant operator to produce a high-quality axion was found, which is made up of Higgs fields for the intermediate symmetry breaking.
In the minimal ${\rm SU}(6)$ GUT, it already possesses a global DRS symmetry as in Eq.~\eqref{eq:DRS}.
With the SUSY extensions, the authors~\cite{Chen:2021haa} found a dimension-$6$ operator that lead to a high-quality axion.
Therefore, it becomes suggestive that the GUTs beyond the minimal versions are likely to solve the PQ-quality problem, with their local and emergent global symmetries.
In the context of GUTs, the PQ-breaking operators can be formed by Higgs fields that develop vacuum expectation values (VEVs) at both the electroweak (EW) scale of $v_{\rm EW}$ and the PQ symmetry-breaking scale of $f_a$.
This further alleviates the PQ quality constraint in Eq.~\eqref{eq:PQquality} even when $d<9$.
For example in the minimal SUSY ${\rm SU}(6)$ GUT~\cite{Chen:2021haa}, such a PQ-breaking operator turns out to be $\mO_{ \bcancel{\rm PQ} }^{d=6}= ( \epsilon_{\alpha\beta} \repb{6_H}^\alpha \repb{6_H}^\beta \rep{15_H} )^2$, with $\alpha=1\,,2$.
The PQ quality constraint can be fulfilled when setting $\langle \repb{6_H}^1 \rangle \sim  \langle \rep{15_H} \rangle \simeq v_{\rm EW}$ and $\langle \repb{6_H}^2 \rangle \simeq f_a$.
A natural question one can raise is whether the PQ quality constraint can be generally satisfied in the non-minimal GUTs with $n_g=3$.
Impressively, we find this generally holds with proper assignment of the PQ charges to the Higgs fields.

The probes of the axion rely on the axion-photon effective coupling of
\beqn
C_{a\gamma\gamma}&=& \frac{E}{ N_{{\rm SU}(3)_c } } -1.92\,.
\eeqn
For the GUTs, there is a universal prediction to the factor $E/ N_{{\rm SU}(3)_c } = 8/3$.
The color anomaly factor of $N_{{\rm SU}(3)_c}$ relates the axion decay constant with the associate symmetry-breaking scale as $v_{\rm SB}= |2 N_{{\rm SU}(3)_c} |\, f_a$, and also determines the domain wall number as $N_{\rm DW} = 2 N_{{\rm SU}(3)_c}$.
In practice, one does not need to derive the factor by analyzing the symmetry breaking patterns.
Instead, this can be obtained by using the 't Hooft anomaly matching condition~\cite{tHooft:1979rat} of
\beqn\label{eq:colorAnom}
N_{{\rm SU}(3)_c} &=& N_{ {\rm SU}(N) } = \sum_{\mathbf F} T( \mR_{\mathbf F} ) {\rm PQ}( \mR_{\mathbf F} )\,.
\eeqn
Notice that in our current study, the physical axion does not arise at the GUT scale of $\sim 10^{16}\,{\rm GeV}$.
Instead, it arises from the phases of Higgs fields that are responsible for the intermediate symmetry breaking scale, with necessary orthogonality conditions imposed.
One such example can be found in the minimal SUSY ${\rm SU}(6)$ GUT~\cite{Chen:2021haa}.
We focus on the PQ-breaking operators in the non-minimal GUTs, while the constructions of the physical axion in the specific GUT model will be left for future work.

\section{The results}
\label{sec:results}

In this section, we obtain our results of the ${\rm SU}(N)$ GUTs that lead to $n_g=3$.
Examples include ${\rm SU}(7)$, ${\rm SU}(8)$, and ${\rm SU}(9)$ groups, where the AF can be achieved.
We also find that the higher groups of ${\rm SU}(10)$ and ${\rm SU}(11)$ with their minimal irreps cannot achieve the AF condition.
For each case, we also look for the possible gauge-invariant and PQ-breaking operators.
With proper PQ charge assignment at the GUT scale, we show that the PQ quality problem can be generally avoided in each model.

\subsection{The ${\rm SU}(7)$}
\label{subsec:SU7}

For the ${\rm SU}(7)$ group, the anomaly vector in Eq.~\eqref{eq:anomaly_vector} reads
\beqn\label{eq:anomaly_SU7}
\vec A&=& (0\,, 1\,, 3\,, 2\,, -2\,, -3 \,, -1)\,.
\eeqn
The decompositions of the ${\rm SU}(7)$ irreps under the ${\rm SU}(5)$ are the following
\beqs
\beqn
\rep{7}&=& \[ 7\,,1 \] = 2\times \[5\,,0 \] \oplus \[5\,,1 \] \,, \\
\rep{21}&=& \[ 7\,,2 \] =  \[5\,,0 \] \oplus 2 \times \[5\,,1 \] \oplus \[5\,,2 \] \,,\\
\rep{35}&=& \[ 7\,,3 \] = \[5\,,1 \] \oplus 2 \times \[5\,,2 \] \oplus \[5\,,3 \]\,.
\eeqn
\eeqs
There are two possibilities for $n_g=3$, namely,
\beqs\label{eqs:fermions_SU7}
\beqn
\{ f_L \}_{ {\rm SU}(7) }^{\rm A}&=& 2\times  \[ 7\,,2 \] \oplus  \[ 7\,,3 \] \oplus 8\times \[ 7\,,6 \] \non
&=&  2 \times \rep{21_F} \oplus \rep{35_F} \oplus  8\times \repb{7_F}  \,,\quad {\rm dim}_{ {\bf F}}= 133 \,,\non
\mG_{\rm global}^{\rm A}&=&\Big[  {\rm SU}(8)_{ \repb{7_F} } \otimes {\rm U}(1)_{\rm PQ} \Big] \otimes \Big[ {\rm SU}(2) \otimes {\rm U}(1)^\prime \Big] \,,\\
\{ f_L \}_{ {\rm SU}(7) }^{\rm B}&=& \[ 7\,,2 \]  \oplus 2\times \[ 7\,,3 \]  \oplus 7\times \[ 7\,,6 \] \non
&=&\rep{21_F}  \oplus  2\times \rep{35_F} \oplus 7\times \repb{7_F}  \,,\quad {\rm dim}_{ {\bf F}}= 140 \,, \non
\mG_{\rm global}^{\rm B}&=&\Big[ {\rm SU}(7)_{ \repb{7_F}} \otimes {\rm U}(1)_{\rm PQ} \Big] \otimes \Big[ {\rm SU}(2) \otimes {\rm U}(1)^\prime \Big] \,.
\eeqn
\eeqs
Since the number of fermions in two cases only differ by less than $10$, we determine to consider both possibilities.
Note in passing, a recent study~\cite{Everett:2021mef} suggests that the ${\rm SU}(7)$ model can be suppress the proton decay with the proper embedding of the SM fermions.

The Higgs sector of two ${\rm SU}(7)$ models is determined by the fermions and the global symmetries in Eq.~\eqref{eqs:fermions_SU7} as follows
\beqs\label{eq:Higgs_SU7}
\beqn
\{ H \}_{ {\rm SU}(7) }^{\rm A}&=& 8\times \repb{21_H} \oplus \rep{7_H} \oplus 2\times \rep{21_H}  \oplus \rep{35_H} \Big[ \oplus \rep{48_H} \Big] \,,\\
\{ H \}_{ {\rm SU}(7) }^{\rm B}&=& 7\times \repb{ 7_H} \oplus \rep{7_H}  \oplus 2\times \rep{21_H} \oplus \rep{35_H} \Big[ \oplus \rep{48_H} \Big]\,.
\eeqn
\eeqs
Here and below, we use the square brackets to denote the real adjoint Higgs fields for the GUT scale symmetry breaking.
By using the fermions and Higgs fields in Eqs.~\eqref{eqs:fermions_SU7} and \eqref{eq:Higgs_SU7}, we find that $b_1^A= - 5$ and $b_1^B=-\frac{55}{6} $.
Thus both the ${\rm SU}(7)$-A and the ${\rm SU}(7)$-B model are asymptotically free.
The gauge-invariant Yukawa couplings are
\beqs\label{eqs:Yukawa_SU7}
\beqn
-\mL_Y^{\rm A}&=&\sum_{\rho=1}^8 \repb{7_F^\rho } \rep{35_F} \repb{21_{H\,\rho} } + \sum_{\dot \rho=1\,,2}   \rep{21_F^{\dot\rho}} \rep{35_F} \rep{21_{H\,{\dot\rho}} }  \non
& +& \epsilon_{ \dot \rho \dot \sigma} \rep{21_F}^{\dot \rho} \rep{21_F}^{\dot \sigma} \rep{35_H} + \rep{ 35_F} \rep{ 35_F} \rep{7_H} + H.c. \,,\\
-\mL_Y^{\rm B}&=& \sum_{\rho=1}^7 \repb{7_F^\rho } \rep{21_F} \repb{7_{H\,\rho} } + \sum_{\dot \rho=1\,,2}  \rep{35_F^{\dot\rho}} \rep{21_F} \rep{21_{H\,{\dot\rho}} }  \non
& +&  \epsilon_{ \dot \rho \dot \sigma} \rep{35_F}^{\dot \rho} \rep{35_F}^{\dot \sigma} \rep{ 7_H} + \rep{ 21_F} \rep{ 21_F} \rep{ 35_H} + H.c. \,.
\eeqn
\eeqs
%

\begin{table}[htp]
\begin{center}
\begin{tabular}{c|ccc|ccccc}
\hline\hline
 ${\rm SU}(7)$-A &  $ \repb{7_F}$ & $\rep{21_F}$ & $\rep{35_F}$ &  $ \repb{21_H}$ & $\rep{ 7_H}$ & $\rep{21_H}$ & $\repb{35_H}$ & $\rep{35_H}$   \\
\hline
PQ charges & $ 1$ & $0$ & $0$ & $-1$ & $0$  & $0$ &  $0$ &  $0$  \\
${\rm SU}(8)_{ \repb{7_F} }$ &  $\Box$  &  $1$ & $1$  &  $\overline\Box$  & $1$  & $1$  & $1$ & $1$  \\
${\rm SU}(2)$ &  $1$  &  $\Box$ &  $1$ &  $1$  & $1$  & $\overline\Box$ &  $\overline\Box$  & $1$ \\
\hline\hline
${\rm SU}(7)$-B &  $ \repb{7_F}$ & $\rep{35_F}$ & $\rep{ 21_F}$ &  $ \repb{7_H}$ & $\rep{ 35_H}$ & $\repb{21_H}$ & $\rep{21_H}$ & $\rep{7_H}$   \\ 
\hline
PQ charges & $ 1$ & $ 0$ & $0$ & $-1$ & $0$  & $0$ &  $0$ &  $0$ \\
${\rm SU}(7)_{ \repb{7_F} }$ &  $\Box$  &  $1$ & $1$  &  $\overline\Box$  & $1$  & $1$  & $1$ & $1$  \\
${\rm SU}(2)$ &  $1$  &  $\Box$ &  $1$ &  $1$  & $1$  & $\overline\Box$ &  $\overline\Box$ & $1$ \\
\hline\hline
\end{tabular}
\end{center}
\caption{The PQ charge assignments and the representations of the fermions and Higgs fields under their global symmetries for two $\gSU(7)$ unification models.}
\label{tab:SU7PQ}
\end{table}%

We assign the PQ charges for all ${\rm SU}(7)$ fermions and Higgs fields in Tab.~\ref{tab:SU7PQ}.
The PQ charges cannot be uniquely determined from the PQ neutrality of the Yukawa couplings~\eqref{eqs:Yukawa_SU7}.
Therefore, we assign the PQ charges by removing the possible dangerous PQ-breaking operators with low mass dimensions.
In the ${\rm SU}(7)$-A, one may assign ${\rm PQ}( \rep{21_F})=q_1$ and ${\rm PQ}( \rep{35_F})=q_2$.
Accordingly, it is easy to find two following PQ-breaking operators
\beqn
&& \mO_{ \bcancel{\rm PQ} }^{d=5}  = \epsilon^{ \dot \rho \dot \sigma} \rep{21_{H\, \dot\rho}} \rep{21_{H\, \dot\sigma}} \rep{7_H}^3 \,, \quad \Delta{\rm PQ} = -2q_1 - 8 q_2 \non
&& \mO_{ \bcancel{\rm PQ} }^{d=3}  = \epsilon^{ \dot \rho \dot \sigma} \rep{21_{H\, \dot\rho} } \repb{35_{H\,\dot \sigma}} \rep{7_H}\,, \quad \Delta {\rm PQ} = -2 q_1 -4 q_2\,.
\eeqn
These two operators would better be PQ-neutral due to their mass dimensions, and this leads to $q_1=q_2=0$.
Similarly in the ${\rm SU}(7)$-B and larger groups below, we find the corresponding PQ charge assignments.

The color and electromagnetic anomaly factors, and domain wall numbers according to Eq.~\eqref{eq:colorAnom} are given by
\beqs
\beqn
{\rm SU}(7)-{\rm A}~&:&~N_{ {\rm SU}(3)_c }= 4\,, \quad E= \frac{32}{3} \,, \quad N_{\rm DW}=8 \,,\\
{\rm SU}(7)-{\rm B}~&:&~ N_{ {\rm SU}(3)_c }=\frac{7}{2} \,, \quad E= \frac{28}{3}\,, \quad  N_{\rm DW}=7 \,,
\eeqn
\eeqs
for two models.
The leading gauge-invariant PQ-breaking operators become~\footnote{Throughout the context, we use the square brackets to anti-symmetrize the indices.}
\beqs
\beqn
{\rm SU}(7)-{\rm A}~&:&~ \mO_{ \Delta{\rm PQ}=-8 }^{d=10}  = ( \repb{21_{H} }_\rho )^8\cdot \rep{7_H}^2 = \epsilon^{A_1\, ...\, A_7 }   \epsilon^{B_1\, ...\, B_7 } \epsilon^{\rho_1\,...\, \rho_8 }  \non
&& ( \repb{21_H}_{\rho_1} )_{\[A_1 B_1\] } ... ( \repb{21_H}_{\rho_8} )_{\[A_8 B_8\] } \rep{7_H}^{ [ A_8 } \rep{7_H}^{B_8 ] } \,, \\
{\rm SU}(7)-{\rm B}~&:&~  \mO_{ \Delta{\rm PQ}=-7 }^{d=7}  = ( \repb{7_H}_\rho )^7  \non
&=& \epsilon^{A_1\,...\, A_7 } \epsilon^{ \rho_1\,...\, \rho_7 } ( \repb{7_H}_{\rho_1} )_{A_1} ... ( \repb{7_H}_{\rho_7} )_{A_7}    \,.
\eeqn
\eeqs
For the leading PQ-breaking operator in the ${\rm SU}(7)$-A model, its mass dimension of $10$ can guarantee the PQ-quality constraint even if all Higgs fields of $\repb{21_H}_\rho$ and $\rep{7_H}$ develop their VEVs at $\sim f_a$.
For the ${\rm SU}(7)$-B model, the leading PQ-breaking operator is of mass dimension $7$.
In order to have a consistent axion decay constant of $f_a\gtrsim 10^8\,{\rm GeV}$, it is necessary that one of the $\repb{7_H}_\rho$ develops its VEV for the electroweak symmetry breaking (EWSB).

\subsection{The ${\rm SU}(8)$}
\label{subsec:SU8}

For the ${\rm SU}(8)$ group, the anomaly vector in Eq.~\eqref{eq:anomaly_vector} reads
\beqn\label{eq:anomaly_SU8}
\vec A&=& (0\,, 1\,, 4\,, 5\,,0\,, -5\,, -4 \,, -1)\,.
\eeqn
The decompositions of the ${\rm SU}(8)$ irreps under the ${\rm SU}(5)$ are the following
\beqs
\beqn
\rep{8}&=& \[ 8\,,1\] = 3\times \[ 5\,,0 \] \oplus \[ 5\,,1\] \,,\\
\rep{28}&=& \[ 8\,,2 \] = 3\times \[5\,,0 \] \oplus 3 \times \[5\,,1 \] \oplus \[5\,,2 \] \,,\\
\rep{56} &=& \[ 8\,,3 \] = \[5\,,0 \] \oplus 3\times \[5\,,1 \] \oplus 3 \times \[5\,,2 \] \oplus \[5\,,3 \]\,.
\eeqn
\eeqs
The possibility for $n_g=3$ with the minimal anomaly-free fermion content is given by
\beqn\label{eq:fermions_SU8}
\{ f_L \}_{ {\rm SU}(8) }&=&  \[ 8\,,2 \]  \oplus  \[ 8\,,3 \] \oplus 9\times \[ 8\,,7 \] = \rep{28_F} \oplus \rep{56_F} \oplus 9\times \repb{8_F} \,,\quad {\rm dim}_{ {\bf F}}= 156 \,, \non
\mG_{\rm global}&=& {\rm SU}(9)_{ \repb{8_F} } \otimes {\rm U}(1)_{\rm PQ} \,.
\eeqn

\begin{table}[htp]
\begin{center}
\begin{tabular}{c|ccc|ccccc}
\hline
 &  $9\times \repb{8_F}$ & $\rep{28_F}$ & $\rep{56_F}$ & $9\times \repb{8_H}$ & $9\times \repb{28_H}$ & $\rep{28_H}$ & $\rep{56_H}$ & $\rep{70_H}$  \\
\hline
PQ charges & $ 1$ & $ 0$ & $0$ & $-1$ & $0$  & $0$ & $0$ & $0$  \\
${\rm SU}(9)_{ \repb{8_F} }$ &  $\Box$  &  $1$  & $1$ & $\overline\Box$  & $1$ & $1$ & $1$ & $1$   \\
\hline
\end{tabular}
\end{center}
\caption{The PQ charge assignments and the ${\rm SU}(9)_F$ representations of the fermions and Higgs fields in the minimal $\gSU(8)$ unification model.}
\label{tab:SU8PQ}
\end{table}%

%
The Higgs sector of the ${\rm SU}(8)$ is determined by the fermions and the global symmetries in Eq.~\eqref{eq:fermions_SU8} as follows~\footnote{In the ${\rm SU}(8)$, the $70$-dimensional irrep is self-conjugate.}
\beqn\label{eq:Higgs_SU8}
\{ H \}_{ {\rm SU}(8) }&=& 9\times \repb{8_H } \oplus 9\times \repb{28_H} \oplus \rep{28_H}\oplus \rep{56_H}  \oplus \rep{70_H} \Big[ \oplus \rep{63_H} \Big] \,.
\eeqn
By using the fermions and Higgs fields in Eqs.~\eqref{eq:fermions_SU8} and \eqref{eq:Higgs_SU8}, we find that $b_1=-\frac{2}{3}<0$.
Thus the minimal ${\rm SU}(8)$ model is asymptotically free.
The gauge-invariant Yukawa couplings are
\beqn\label{eq:Yukawa_SU8}
-\mL_Y&=&\sum_{\rho=1}^9 \Big(  \repb{8_F}^\rho \rep{28_F}  \repb{8_{H}}_\rho + \repb{8_F}^\rho \rep{56_F}  \repb{28_{H}}_\rho \Big) \non
&+& \rep{56_F} \rep{56_F} \rep{28_H}  +\rep{28_F} \rep{56_F} \rep{56_H}+   \rep{28_F} \rep{28_F} \rep{70_H} + H.c.\,,
\eeqn
We assign PQ charges for all ${\rm SU}(8)$ fields in Tab.~\ref{tab:SU8PQ}, with the same argument in the ${\rm SU}(7)$ models.
The color and electromagnetic anomaly factors, and domain wall numbers according to Eq.~\eqref{eq:colorAnom} are given by
\beqn
{\rm SU}(8)~&:&~N_{ {\rm SU}(3)_c }= \frac{9}{2}\,, \quad E= 12 \,, \quad N_{\rm DW}=9 \,.
\eeqn
The leading PQ-breaking operators in the ${\rm SU}(8)$ model is
\beqn
{\rm SU}(8)~&:&~ \mO_{ \Delta{\rm PQ}=-9 }^{d=12}  = ( \repb{8_{H\, \rho} } )^9\cdot \rep{28_H} \cdot \rep{56_H} \cdot \rep{70_H} \non
&=& \epsilon^{A_1\, ...\, A_8} \epsilon_{B_1\, ...\, B_8} \epsilon^{\rho_1\, ...\, \rho_8\, \delta} ( \repb{8_H}_{\rho_1} )_{A_1} ... ( \repb{8_{H}}_{\rho_8} )_{A_8}  ( \repb{8_{H}}_{\delta} )_{B_0} \non
&& ( \rep{28_H} )^{ [ B_0\, B_1] }  ( \rep{56_H} )^{ [ B_2\, B_3\, B_4] }  ( \rep{70_H} )^{ [ B_5\, B_6\, B_7 \, B_8] }  \,.
\eeqn

\subsection{The ${\rm SU}(9)$}
\label{subsec:SU9}

For the ${\rm SU}(9)$ group, the anomaly vector in Eq.~\eqref{eq:anomaly_vector} reads
\beqn\label{eq:anomaly_SU9}
\vec A&=& (0\,, 1\,, 5 \,, 9 \,, 5 \,, -5\,, -9\,, -5\,, -1)\,.
\eeqn
The decompositions of the ${\rm SU}(9)$ irreps under the ${\rm SU}(5)$ are the following
\beqs
\beqn
\rep{9}&=& \[ 9\,,1 \] = 4\times \[ 5\,,0 \] \oplus \[ 5\,,1 \] \,,\\
\rep{36}&=& \[ 9\,,2 \]= 6\times \[ 5\,,0 \] \oplus 4\times \[ 5\,,1 \] \oplus \[ 5\,,2 \] \,,\\
\rep{84 }&=& \[ 9\,,3 \]= 4\times \[ 5\,,0 \] \oplus 6\times \[ 5\,,1 \] \oplus 4\times \[ 5\,,2 \] \oplus \[ 5\,,3 \] \,,\\
\rep{126}&=& \[ 9\,,4 \] = \[ 5\,,0 \] \oplus 4\times \[ 5\,,1 \] \oplus 6\times \[ 5\,,2 \] \oplus 4\times \[ 5\,,3 \] \oplus \[ 5\,,4 \]  \,.
\eeqn
\eeqs
The possibility for $n_g=3$ with the minimal anomaly-free fermion content is given by
\beqn\label{eq:fermions_SU9}
\{ f_L \}_{ {\rm SU}(9) }&=&  \[ 9\,,3 \] \oplus 9\times \[ 9\,,8 \] =  \rep{84_F} \oplus 9\times \repb{9_F}  \,,\quad {\rm dim}_{ {\bf F}}= 165 \,,\non
\mG_{\rm global}&=& {\rm SU}(9)_{ \repb{9_F} } \otimes {\rm U}(1)_{\rm PQ} \,.
\eeqn
Another possibility with more fermions of $\[ 9\,,2 \] \oplus \[ 9\,,4 \] \oplus 10\times \[ 9\,,8 \] $ will not be considered here.
An even larger fermion contents of $ 2\times \[ 9\,,2 \] \oplus 2\times \[ 9\,,4 \] \oplus  \[ 9\,,6 \]  \oplus 11\times \[ 9\,,8 \]   $ were previously mentioned in Ref.~\cite{Barr:1979xt}.

\begin{table}[htp]
\begin{center}
\begin{tabular}{c|cc|ccc}
\hline
 &  $9\times \repb{9_F}$ & $\rep{84_F}$ & $9\times \repb{36_H}$ & $\rep{84_H}$  \\
\hline
PQ charges & $ 1$ & $ 0$ & $-1$ & $0$    \\
${\rm SU}(9)_{ \repb{9_F} }$ &  $\Box$  &  $1$  &  $\overline\Box$  & $1$   \\
\hline
\end{tabular}
\end{center}
\caption{The PQ charge assignments and the ${\rm SU}(9)_F$ representations of the fermions and Higgs fields in the minimal $\gSU(9)$ unification model.}
\label{tab:SU9PQ}
\end{table}%

%
The Higgs sector of the ${\rm SU}(9)$ is determined by the fermions and the global symmetries in Eq.~\eqref{eq:fermions_SU9} as follows
\beqn\label{eq:Higgs_SU9}
\{ H \}_{ {\rm SU}(9) }&=&  9\times \repb{36_H} \oplus \rep{84_H} \Big[ \oplus \rep{80_H} \Big] \,.
\eeqn
By using the fermions and Higgs fields in Eqs.~\eqref{eq:fermions_SU9} and \eqref{eq:Higgs_SU9}, we find that $b_1=-\frac{15}{2}<0$.
Thus the minimal ${\rm SU}(9)$ model is asymptotically free.
The gauge-invariant Yukawa couplings are
\beqn\label{eq:Yukawa_SU9}
-\mL_Y&=& \repb{9_F}^\rho \rep{84_F}  \repb{36_{H}}_\rho + \rep{84_F} \rep{84_F} \rep{84_H} + H.c.\,,
\eeqn
and we assign the PQ charges in Tab.~\ref{tab:SU9PQ}.
Naively, the PQ charge of the $\rep{84_H}$ can be arbitrary according to the Yukawa couplings~\eqref{eq:Yukawa_SU9}.
It turns out a dimension-$3$ PQ-breaking operator of $\mO_{ \bcancel{\rm PQ} }^{d=3}=( \rep{84_H})^3= \epsilon_{A_1 B_1 C_1 A_2 B_2 C_2 A_3 B_3 C_3} (\rep{84_H})^{[A_1 B_1 C_1 ]} (\rep{84_H})^{ [ A_2 B_2 C_2 ]} (\rep{84_H})^{ [ A_3 B_3 C_3 ]}$ can arise, which is dangerous from the dimensional counting.
Therefore, we determine that ${\rm PQ}(\rep{84_H})=0$.
The color and electromagnetic anomaly factors, and domain wall numbers according to Eq.~\eqref{eq:colorAnom} are given by
\beqn
{\rm SU}(9)~&:&~N_{ {\rm SU}(3)_c }= \frac{9}{2}\,, \quad E= 12 \,, \quad N_{\rm DW}=9 \,.
\eeqn
The leading dimension-$9$ PQ-breaking operator in the ${\rm SU}(9)$ model is
\beqn\label{eq:PQ_SU9}
 \mO_{ \Delta{\rm PQ}=-9 }^{d=9} &=& ( \repb{36_H})^9= \epsilon^{A_1\,...\, A_9} \epsilon^{B_1\,...\, B_9} \epsilon^{\rho_1\,...\, \rho_9} ( \repb{36_H}_{\rho_1} )_{[ A_1 B_1 ] }\,...\, ( \repb{36_H}_{\rho_9} )_{[ A_9 B_9] } \,.
\eeqn
According to the dimension counting in Ref.~\cite{Kamionkowski:1992mf}, this is likely to produce a high-quality axion.

\subsection{The ${\rm SU}(10)$}

For the ${\rm SU}(10)$ group, the anomaly vector in Eq.~\eqref{eq:anomaly_vector} reads
\beqn\label{eq:anomaly_SU10}
\vec A&=& (0\,, 1\,, 6 \,,  14 \,, 14\,, 0\,, -14\,, -14 \,, -6 \,, -1)\,.
\eeqn
The decompositions of the ${\rm SU}(10)$ irreps under the ${\rm SU}(5)$ are the following
\beqs
\beqn
\rep{10}&=& \[10\,,1 \] = 5\times \[ 5\,,0 \] \oplus \[ 5\,,1\] \,,\\
\rep{45}&=& \[10\,,2 \] = 10\times \[ 5\,,0 \] \oplus 5\times \[ 5\,,1\] \oplus \[ 5\,,2\] \,,\\
\rep{120}&=& \[10\,,3 \] = 10\times \[ 5\,,0 \] \oplus 10\times \[ 5\,,1\] \oplus 5\times \[ 5\,,2\] \oplus \[ 5\,,3\]  \,,\\
\rep{210}&=& \[10\,,4 \] = 5\times \[ 5\,,0 \] \oplus 10\times \[ 5\,,1\] \oplus 10\times \[ 5\,,2\] \oplus 5\times \[ 5\,,3\] \oplus \[ 5\,,4\]  \,.
\eeqn
\eeqs
The possibility for $n_g=3$ with the minimal anomaly-free fermion content is given by
\beqn\label{eq:fermions_SU10}
\{ f_L \}_{ {\rm SU}(10) }&=& \[ 10\,,3 \]  \oplus  \[ 10\,,8 \] \oplus 8\times \[ 10\,,9 \] \non
&=& \rep{120_F}\oplus \repb{45_F} \oplus 8\times \repb{10_F} \,,\quad {\rm dim}_{ {\bf F}}= 245 \,,\non
\mG_{\rm global}&=& {\rm SU}(8)_{ \repb{10_F} } \otimes {\rm U}(1)_{\rm PQ} \,.
\eeqn

\begin{table}[htp]
\begin{center}
\begin{tabular}{c|ccc|ccccc}
\hline
 &  $8\times \repb{10_F}$ & $\repb{ 45_F}$ & $\rep{ 120_F}$  & $8\times \repb{ 45_H}$ & $8\times \rep{ 120_H}$ & $\repb{ 10_H}$ & $\rep{ 210_H}$  \\
\hline
PQ charges & $ 1$ & $ 0$  & $ 0$  & $-1$ & $-1$  & $0$  & $0$  \\
${\rm SU}(8)_{ \repb{10_F} }$ &  $\Box$  &  $1$  &  $1$  &  $\overline\Box$ &  $\overline\Box$ & $1$  & $1$  \\
\hline
\end{tabular}
\end{center}
\caption{The PQ charge assignments and the ${\rm SU}(8)_{ \repb{10_F} }$ representations of the fermions and Higgs fields in the minimal $\gSU(10)$ unification model.}
\label{tab:SU10PQ}
\end{table}%

%
The Higgs sector of the ${\rm SU}(10)$ is determined by the fermions and the global symmetries in Eq.~\eqref{eq:fermions_SU10} as follows
\beqn\label{eq:Higgs_SU10}
\{ H \}_{ {\rm SU}(10) }&=&  8\times \repb{45_H} \oplus 8\times \rep{120_H} \oplus \repb{10_H} \oplus \rep{210_H} \Big[ \oplus \rep{99_H} \Big] \,.
\eeqn
By using the fermions and Higgs fields in Eqs.~\eqref{eq:fermions_SU10} and \eqref{eq:Higgs_SU10}, we find that $b_1=\frac{223}{6}$.
Thus the minimal ${\rm SU}(10)$ model is not asymptotically free.
The gauge-invariant Yukawa couplings are
\beqn\label{eq:Yukawa_SU10}
-\mL_Y&=& \repb{10_F}^\rho \rep{120_F}  \repb{ 45_{H}}_\rho + \repb{10_F}^\rho \repb{45_F}  \rep{ 120_{H}}_\rho \non
&+& \Big( \repb{ 45_F} \repb{ 45_F}+ \rep{120_F} \rep{120_F} \Big) \rep{210_H} + \repb{ 45_F} \rep{ 120_F}  \repb{10_H}  +  H.c.\,,
\eeqn
and we assign the PQ charges in Tab.~\ref{tab:SU10PQ}.
It turns out that three dimension-5 PQ-breaking operators of $\mO_{ \bcancel{\rm PQ} }^{d=5 }= ( \repb{10_H} )^4 \rep{210_H}$, $\mO_{ \bcancel{\rm PQ} }^{d=5 }= ( \repb{10_H} )^2 (\rep{210_H})^3$, and $\mO_{ \bcancel{\rm PQ} }^{d=5 }= ( \rep{210_H} )^5$ can arise if ${\rm PQ}( \repb{10_H} ) \neq 0$ or ${\rm PQ}( \rep{210_H} ) \neq 0$.
Therefore, we determine that ${\rm PQ}( \repb{10_H} ) = 0$ and ${\rm PQ}( \rep{210_H} ) = 0$.
The color and electromagnetic anomaly factor, and domain wall numbers according to Eq.~\eqref{eq:colorAnom} are given by
\beqn
{\rm SU}(10)~&:&~N_{ {\rm SU}(3)_c }= 4\,, \quad E= \frac{32}{3} \,, \quad N_{\rm DW}= 8 \,.
\eeqn
Two dimension-12 PQ-breaking operators in the ${\rm SU}(10)$ model are expressed as follows
\beqs\label{eq:PQ_SU10}
\beqn
\mO_{ \Delta{\rm PQ}=-8 }^{d=12} &=& ( \repb{ 45_H})^8  ( \repb{10_H})^2 ( \rep{210_H})^2  \non
&=& \epsilon^{ A_1\, B_1\, ... A_5\, B_5  }    \epsilon^{ \rho_1\,... \rho_8  } (  \repb{45_H}_{\rho_1}  )_{[ A_1 B_1 ] } ... (  \repb{45_H}_{\rho_5}  )_{[ A_5 B_5 ] } \non
&& (  \repb{45_H}_{\rho_6}  )_{[ A_6 B_6 ] } (  \repb{45_H}_{\rho_7}  )_{[ A_7 B_7 ] } (  \repb{45_H}_{\rho_8 }  )_{[ A_8 B_8 ] } (  \repb{10_H}  )_{C_1}  (  \repb{10_H}  )_{C_2} \non
&& (\rep{210_H}  )^{ [ A_6 A_7 A_8 C_1 ] }  (\rep{210_H}  )^{ [ B_6 B_7 B_8 C_2 ] }  \,,\\
\mO_{ \Delta{\rm PQ}=-8 }^{d=12} &=& ( \rep{120_H})^8  ( \repb{10_H})^2 ( \rep{210_H})^2 \non
&=& \epsilon^{ A_1\, ... \, A_8 A_9 A_{10} }  \epsilon^{ B_1\,... \, B_8 B_9 B_{10}   } \epsilon^{ C_1\,... \, C_8 C_9 C_{10}  }  \epsilon^{ \rho_1\,... \rho_8  } \non
&&( \rep{120_H}_{\rho_1} )^{ [ A_1 B_1 C_1 ] } ... ( \rep{120_H}_{\rho_8} )^{ [ A_8 B_8 C_8 ] } (  \repb{10_H}  )_{D_1}  (  \repb{10_H}  )_{D_2} \non
&&  (\rep{210_H}  )^{ [  D_1 A_9 B_9 C_9 ] }  (\rep{210_H}  )^{ [ D_2 A_{10} B_{10} C_{10} ] }   \,.
\eeqn
\eeqs

\subsection{The ${\rm SU}(11)$}
\label{sec:SU11}

For the ${\rm SU}(11)$ group, the anomaly vector in Eq.~\eqref{eq:anomaly_vector} reads
\beqn\label{eq:anomaly_SU11}
\vec A&=& (0\,, 1\,, 7 \,, 20\,, 28\,, 14\,, -14\,, -28 \,, -20\,, -7 \,, -1)\,.
\eeqn
The decompositions of the ${\rm SU}(11)$ irreps under the ${\rm SU}(5)$ are the following
\beqs
\beqn
\rep{11}&=& \[11\,,1 \]= 6\times \[5\,,0 \] \oplus \[5\,,1 \] \,,\\
\rep{55}&=& \[11\,,2 \] = 15\times \[5\,,0 \] \oplus 6\times \[5\,,1 \] \oplus  \[5\,,2 \] \,,\\
\rep{165}&=& \[11\,,3 \] = 20 \times \[5\,,0 \] \oplus 15\times \[5\,,1 \] \oplus 6\times \[5\,,2 \] \oplus \[5\,,3 \]  \,,\\
\rep{330}&=& \[11\,,4 \] = 15\times \[5\,,0 \] \oplus 20\times \[5\,,1 \] \oplus 15\times \[5\,,2 \] \oplus 6\times \[5\,,3 \] \oplus \[5\,,4\] \,,\\
\rep{462}&=& \[11\,,5 \] = 7\times \[5\,,0 \] \oplus 15\times \[5\,,1 \] \oplus 20\times  \[5\,,2 \] \oplus 15\times  \[5\,,3 \] \oplus 6\times \[5\,,4 \]    \,.
\eeqn
\eeqs
The possibility for $n_g=3$ with the minimal anomaly-free fermion content is given by
\beqn\label{eq:fermions_SU11}
\{ f_L \}_{ {\rm SU}(11) }&=& \[ 11\,,3 \]  \oplus 2\times  \[ 11\,,9 \]   \oplus 6\times \[ 11\,,10 \] \non
&=& \rep{165_F} \oplus 2 \times \repb{55_F} \oplus 6 \times \repb{11_F} \,,\quad {\rm dim}_{ {\bf F}}= 341 \,,\non
\mG_{\rm global}&=& \Big[ {\rm SU}(6)_{ \repb{11_F} } \otimes {\rm U}(1)_{\rm PQ} \Big] \otimes \Big[ {\rm SU}(2) \otimes {\rm U}(1)^\prime \Big] \,.
\eeqn

\begin{table}[htp]
\begin{center}
\begin{tabular}{c|ccc|ccccc}
\hline
 &  $6\times \repb{11_F}$ & $2\times \repb{ 55_F}$ & $\rep{ 165_F}$  & $6\times \repb{ 55_H}$ & $2\times \repb{ 11_H}$ & $\rep{ 330_H}$ & $\rep{ 462_H}$  \\
\hline
PQ charges & $ 1$ & $ 0$  & $ 0$  & $-1$ & $-1$  & $0$  & $0$  \\
${\rm SU}(6)_{ \repb{11_F} }$ &  $\Box$  &  $1$  &  $1$  &  $\overline\Box$ &  $\overline\Box$ & $1$  & $1$  \\
\hline
\end{tabular}
\end{center}
\caption{The PQ charge assignments and the ${\rm SU}(6)_{ \repb{11_F} }$ representations of the fermions and Higgs fields in the minimal ${\rm SU}(11)$ unification model.}
\label{tab:SU11PQ}
\end{table}%

%
The Higgs sector of the ${\rm SU}(11)$ is determined by the fermions and the global symmetries in Eq.~\eqref{eq:fermions_SU11} as follows
\beqn\label{eq:Higgs_SU11}
\{ H \}_{ {\rm SU}(11) }&=&  6\times \repb{55_H} \oplus 2\times \repb{11_H} \oplus \rep{330_H} \oplus \rep{462_H} \Big[ \oplus \rep{120_H} \Big] \,.
\eeqn
By using the fermions and Higgs fields in Eqs.~\eqref{eq:fermions_SU11} and \eqref{eq:Higgs_SU11}, we find that $b_1=\frac{ 83}{3}$.
The gauge-invariant Yukawa couplings are
\beqn\label{eqs:Yukawa_SU11}
-\mL_Y&=&\sum_{\rho=1}^6 \repb{11_F}^\rho \rep{ 165_F} \repb{ 55_{H}}_\rho + \sum_{\dot \rho=1\,,2}   \repb{ 55_F}^{\dot\rho} \rep{165_F} \repb{ 11_{H}}_{\dot\rho}  \non
& +& \epsilon_{ \dot \rho \dot \sigma} \repb{ 55_F}^{\dot \rho} \repb{55_F}^{\dot \sigma} \rep{330_H} + \rep{ 165_F} \rep{ 165_F} \rep{462_H} + H.c. \,.
\eeqn
We assign the PQ charges in Tab.~\ref{tab:SU11PQ}.
It turns out a possible dimension-5 PQ-breaking operator of $\mO_{ \bcancel{\rm PQ} }^{d=5 }= ( \rep{330_H} )^3 (\rep{462_H})^2$ may arise.
For this reason we consider ${\rm PQ}(\rep{330_H})=0$ and ${\rm PQ}(\rep{462_H})=0$.
The color and electromagnetic anomaly factor, and domain wall numbers according to Eq.~\eqref{eq:colorAnom} are given by
\beqn
{\rm SU}(11)~&:&~N_{ {\rm SU}(3)_c }=3 \,, \quad E= 8 \,, \quad N_{\rm DW}= 6 \,.
\eeqn
The leading PQ-breaking operators in the ${\rm SU}(11)$ model is found to be
\beqn\label{eq:PQ_SU11}
\mO_{ \bcancel{\rm PQ} }^{d=9 } &=&  ( \repb{ 55_H})^6  ( \rep{ 330_H})^3 \non
&=&\epsilon_{ A_1 B_1 \, ... A_6 } \epsilon^{ \rho_1\, ... \, \rho_6 } ( \repb{55_H}_{ \rho_1} )_{[ A_1 B_1 ]} ... ( \repb{55_H}_{ \rho_6} )_{[ A_6 C ]} \non
&& \epsilon_{ D_1 E_1 \, ... \, E_3 F_3 }  ( \rep{330_H} )^{ [D_1 E_1 F_1 G_1 ]}  ( \rep{330_H} )^{[ D_2 E_2 F_2 G_2] }  ( \rep{330_H} )^{ [ D_3 E_3 F_3 C ] }  \,.
\eeqn

\section{Conclusions}
\label{sec:conclusion}

\begin{table}[htp]
\begin{center}
\begin{tabular}{c|ccccc}
\hline\hline
 Models & ${\rm dim}_{\mathbf F}$ & $b_1$ & $N_{ \gSU(3)_c}$   & PQ-breaking \\ \hline
 ${\rm SU}(7)$-A &  $133$ & $-5$ &  $4$    & $( \repb{21_{H} }_\rho )^8\cdot ( \rep{7_H})^2$  \\ 
 ${\rm SU}(7)$-B &  $140$ & $-\frac{55}{6}$ &  $7/2$ &    $( \repb{7_H}_\rho )^7$  \\
 ${\rm SU}(8)$ &  $156$ & $-\frac{2}{3}$ &   $9/2$   & $( \repb{8_{H\, \rho} } )^9\cdot \rep{28_H} \cdot \rep{56_H} \cdot \rep{70_H}$   \\ 
 ${\rm SU}(9)$ &  $165$ & $-\frac{15}{2}$ &  $9/2$ &  $( \repb{36_H})^9$  \\
 ${\rm SU}(10)$ &  $245$ & $\frac{223}{6}$ &  $4$ &  $ ( \repb{45_H})^8  ( \repb{10_H})^2  ( \rep{210_H})^2$  \\
 &  &  &  & $( \rep{120_H})^8  ( \repb{10_H})^2  ( \rep{210_H})^2$  \\
 ${\rm SU}(11)$ &  $341$ & $\frac{83}{3}$ &  $3$ &  $( \repb{ 55_H})^6  ( \rep{ 330_H})^3$  \\
\hline\hline
\end{tabular}
\end{center}
\caption{The non-minimal GUTs with $n_g=3$, their one-loop $\beta$ coefficients, the color anomaly factors, and the PQ-breaking operators.}
\label{tab:summary}
\end{table}%

%
%
We have studied the set of non-minimal GUTs that can lead to the observed three generational SM fermions according to Georgi's counting.
With the origin of the generational structure, these models themselves can be appealing to answer the most puzzling question of the SM fermion mass hierarchies.
Our results suggest four such models that achieve the AF property at short distances, and two more that may be considered with further studies.
The results are summarized in Tab.~\ref{tab:summary}.
The other important feature of the non-minimal GUTs in our study comes from their global symmetries, which was also previously noted in the ${\rm SU}(6)$ model~\cite{Dvali:1993yf,Chen:2021haa}.
Though the ${\rm SU}(6)$ model enjoys a global DRS symmetry, it turns out the SUSY extension was inevitable in order to produce a high-quality PQ symmetry.
In six non-minimal GUTs of the current study, the sizes of the PQ-breaking effects due to the quantum gravity are generally under better control due to the gauge symmetries and the associated global DRS symmetries.
It is thus reasonable to expect the long-standing PQ quality problem can be avoided in non-minimal GUTs with $n_g=3$, where the emergent global DRS symmetries are general.

Obviously, we expect the following studies to be performed for specific models, which include: (i) the viable symmetry breaking patterns, (ii) SM fermion mass hierarchies and their mixings, (iii) the physical axion mass predictions and the related experimental searches.
A recent study of the ${\rm SU}(6)$ toy model~\cite{Chen:2021zwn} suggests that the bottom quark and tau lepton masses can be naturally suppressed to the top quark mass through the seesaw-like mass matrices with their heavy fermion partners.
Such heavy fermion partners for both the down-type quarks and charged leptons are general in the non-minimal GUTs with $n_g=3$.
Furthermore, the non-minimal GUTs with $n_g=3$ can naturally lead to multiple symmetry breaking scales between the $\Lambda_{\rm GUT}$ and the EW scale.
Altogether, we expect that the observed fermion mass hierarchies among three generations can be realized with the appropriate symmetry breaking pattern in the non-minimal GUT with $n_g=3$.
Above all, one has to analyze the gauge coupling unifications for the viable symmetry breaking patterns and predicts the proton lifetime.
Since our models possess several intermediate scales, this usually requires the two-loop RGEs, together with the matching conditions and mass threshold effects~\cite{Weinberg:1980wa,Hall:1980kf}.

\section*{Acknowledgements}

We would like to thank Luca Di Luzio, Jing Ren, and Ye-Ling Zhou for very useful discussions and communication. 
N.C. would like to thank Tibet University for hospitality when preparing this work.
This work is partially supported by the National Natural Science Foundation of China (under Grant No. 12035008 and No. 11575176).


\end{document}